\documentclass[pra,aps,showpacs]{revtex4}
\usepackage{epsfig}
\usepackage[english]{babel}
\usepackage{latexsym}
\usepackage{graphics}
\usepackage{subfigure}
\begin{document}
\title{Entanglement Concentration of Individual Photon Pairs via Linear Optical Logic}
\author{Chuanwei Zhang}
\begin{abstract}
We propose a scheme for concentrating nonmaximally pure and mixed
polarization-entangled state of individual photon pairs. The scheme uses only
simple linear optical elements and may be feasible within current optical
technology.
\end{abstract}
\address{Department of Physics and Center for Nonlinear Dynamics, The
University of Texas, Austin, Texas 78712-1081}
\maketitle
Entanglement has played a key role in many quantum information processing,
such as quantum computation \cite{divin}, quantum teleportation \cite
{teleportation}, quantum dense coding \cite{dense}, and
entanglement-assisted quantum cryptography \cite{Naik00}. To function
optimally these applications requires maximally entanglement. However,
unwanted coupling to the environment causes the degradation of entanglement
and entanglement concentration \cite{Ben96} is thus essential for quantum
computing. The basic idea of entanglement concentration is to distill some
pairs of particles in highly entangled states from less entangled states
using local quantum operations and two-way classical communications (LOCC) 
\cite{divin}. In practice, there are two fundamentally different types of
concentration protocols: those acting on individual pairs of entangled
particles \cite{Vidal99} and those acting collectively on many pairs \cite
{Linden98}.
In recent years, entanglement concentration using linear optical elements
has received much attention \cite{Pan00,Thew01,Kwiat01}. In the case of
concentration of individual entanglement photon pairs, Thew and Munro
proposed a protocol based on beam splitters with variable transmission
coefficients (VBS) \cite{Thew01} and experimentally, Kwiat \textit{et. al. 
} has implemented individual entanglement concentration using partial
polarizers \cite{Kwiat01}. However, both Thew and Munro's protocol and
Kwiat's experiment included parameters that are difficult to adjust in
practice: Thew and Munro's scheme requires four VBS and the partial
polarizers in Kwiat's experiment must be changed according to the
initial entanglement.

In this paper, we propose a scheme for concentrating nonmaximally pure and
mixed polarization-entangled state of individual photon pairs using linear
optical elements (polarization beam splitter (PBS), half wave plate (HWP),
and quarter wave plate (QWP)) . The scheme uses only Mach-Zehnder
interferometers and a few adjustable polarization rotations (generated by
HWP and QWP) and maybe greatly simplify the experiment.

The crucial part in any individual pairs' entanglement concentration scheme
is the realization of single-qubit local quantum operator, including unitary
rotation and the positive-operator-valued measurement (POVM) \cite{pop}.
Generally, a single-qubit unitary rotation on the polarization of a photon
(or single-qubit polarization rotation (SPR)) has the form $U=\left( 
\begin{array}{ll}
e^{-i\xi }\cos \theta & e^{-i\iota }\sin \theta \\ 
e^{i\iota }\sin \theta & -e^{i\xi }\cos \theta
\end{array}
\right) $ and can be implemented using a wave plate sequence as shown in
Fig. 1a, where two phase shifters provide the phase factors $e^{-i\xi }$ and 
$e^{i\iota }$ and one HWP gives the rotation \cite{sten96}. Consider a
single qubit POVM $M_i$ $(i=1,2)$ satisfying $M_1^{\dagger }M_1+M_2^{\dagger
}M_2=I_2$. They can be represented as $M_1=$diag$\left( \cos \theta ,\cos
\vartheta \right) $, $M_2=$diag$\left( \sin \theta ,\sin \vartheta \right) $
up to some unitary operations, where $I_2$ is the unit operation \cite{proof}%
. By using location of each photon as assistant qubit, $M_i$ can be replaced
by a two-qubit unitary operator 
\begin{equation}
U=\left( 
\begin{array}{cc}
R_y\left( -2\theta \right) & 0 \\ 
0 & R_y\left( -2\vartheta \right)
\end{array}
\right)  \label{1}
\end{equation}
acting on both polarization and location, where we have used basis $\left\{
\left| 0\right\rangle _P\left| 0\right\rangle _L,\left| 0\right\rangle
_P\left| 1\right\rangle _L,\left| 1\right\rangle _P\left| 0\right\rangle
_L,\left| 1\right\rangle _P\left| 1\right\rangle _L\right\} $, $R_y\left(
\theta \right) $ is a rotation by $\theta $ around $\hat{y}$, $\left|
i\right\rangle _P$ and $\left| i\right\rangle _L$ represent the polarization
and location, respectively. It is easy to testify that $U$ has the
decomposition 
\[
U=V_1V_2V_3V_2V_1, 
\]
where $V_1$ is a location controlling polarization NOT gate (by adding $%
\sigma _x$ on location $\left| 1\right\rangle _L$), $V_2$ is a polarization
controlling location NOT gate (by a PBS), and $V_3$ represents a location
controlling polarization unitary rotation that performs polarization
rotation $R_y\left( -2\theta \right) $ ($R_y\left( 2\vartheta \right) $) if
the location is $\left| 0\right\rangle _L$ ($\left| 1\right\rangle _L$).
With the decomposition of $U$, a POVM on the polarization of a photon can be
realized using a Mach-Zehnder interferometer as shown in Fig. 1b. 

\begin{figure}[!t]
\begin{center}
\vspace*{-0.0cm}
\par
\resizebox *{6cm}{2cm}{\includegraphics*{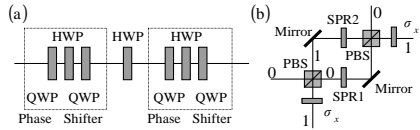}}
\end{center}
\par
\vspace*{-0.0cm}
\caption{Linear optical realizations of arbitrary local single qubit quantum
operator. (a) Single qubit polarization rotation (SPR). (b) Single-qubit
POVM. SPR1 and SPR2 perform operation $R_y\left( -2\theta \right) $ and $%
R_y\left( 2\vartheta \right) $, respectively. }
\label{fig:ele}
\end{figure}

With the linear optics realization of the single qubit local quantum
operator, we can readily study the entanglement concentration of individual
photon pairs. Consider the initial state is an entanglement shared by two
spatially separated subsystem, $A$ and $B$. The qubits used here are
polarization states of the photon with $\left| H\right\rangle $
(Horizontal), $\left| V\right\rangle $ (Vertical) corresponding to the $%
\left| 0\right\rangle _P$, $\left| 1\right\rangle _P$ states above. The
experimental arrangement is described by the schematic plot in Fig. 2. The
left part, including the BBO crystal and quartz decoherers provides the
source of the initial polarization entangled pure \cite{Kwiat95,White99} or
mixed \cite{Kwiat002} states. The entangled photon pair is then incident to
the concentration part. By varying the single-qubit polarization rotation
(SPR), it can perform arbitrary local single-qubit quantum operation and
realize the entanglement concentration. With the prior knowledge of the
initial and final states, SPR $A(B)i$ can be determined and adjusted. The
final SPR in each arm, along with PBS, enable analysis of the polarization
corrections in any basis, allowing tomographic reconstruction of the density
matrix \cite{Kwiat95,White99,Kwiat002,Kwiat01}. In practice, the tomographic
measurements are only performed on the paths with successful entanglement
concentration. In the experiment to concentrate entangled pure states, only
one Mach-Zehnder interferometer is sufficient.

We first consider the behavior of pure states under the protocol. The
concentration is from the partially entangled pure state $\left| \Phi
\right\rangle =\cos \alpha \left| HH\right\rangle +\sin \alpha \left|
VV\right\rangle $ to $\left| \Psi \right\rangle =\cos \beta \left|
HH\right\rangle +\sin \beta \left| VV\right\rangle $, where angles $\alpha
,\beta \in \left[ 0,\pi /4\right] $ and $\alpha <\beta $. The concentration
can be implemented by a POVM $M_1=$diag$\left( \cos \omega ,1\right) $ with
optimal successful probability \cite{Vidal99} 
\begin{equation}
P=\frac{\sin ^2\alpha }{\sin ^2\beta },  \label{2}
\end{equation}
where $\omega =\arccos \left( \tan \alpha /\tan \beta \right) $. This POVM
corresponds to the unitary operation $U=$diag$\left( R_y\left( -2\omega
\right) ,I_2\right) $ on photon $A$ with post-selecting location $\left|
0\right\rangle _L$ as the successful output. In Fig. 2, we set SPR $A2$ to
perform rotation $R_y\left( -2\omega \right) $ and all others are the unit
operation. For example, the only operation for concentrating initial
entanglement $\left| \Phi \right\rangle =\frac{\sqrt{3}}2\left|
HH\right\rangle +\frac 12\left| VV\right\rangle $ to maximal $\left| \Phi
\right\rangle =\frac{\sqrt{2}}2\left| HH\right\rangle +\frac{\sqrt{2}}%
2\left| VV\right\rangle $ is the adjustment of SPR $A2$ to perform
polarization rotation $R_y\left( -2\arccos \left( \frac 1{\sqrt{3}}\right)
\right) $.

Now we turn our attention to the concentration of mixed states. Generally,
an arbitrary bipartite density matrix can be represented as \cite{Schl95} $%
\rho =\left( \sum_{i,j}R_{ij}\sigma _i\otimes \sigma _j\right) /4$, where
the summation extends from $0$ to $3$ with $\sigma _0$ the $2\times 2$
identity matrix and $\sigma _1$, $\sigma _2$, $\sigma _3$ the Pauli spin
matrices, $R_{ij}$ are real and linear parameterizations. Optimal
concentration protocol can be obtained in two cases. If the matrix $R=\left[
R_{ij}\right] $ is diagonalizable by proper orthotropous Lorentz
transformations (POLT), a Bell diagonal mixed state can be extracted with
maximal possible entanglement of formation from the initial mixed state with
nonzero probability \cite{Linden98,Vers00}. If $R$ is not diagonalizable by
POLT, the probability of obtaining Bell-diagonal state is equal to zero.
However, it can still be quasi-distillated \cite{Vers00,Horo98}. Optimal
local quantum operation can be calculated explicitly according to the POLT 
\cite{Vers00}. For both cases, the local quantum operator can be written in
the form 
\begin{equation}
U_A\left( 
\begin{array}{cc}
\cos \theta _A & 0 \\ 
0 & \cos \delta _A
\end{array}
\right) U_A^{^{\prime }}\otimes U_B\left( 
\begin{array}{cc}
\cos \theta _B & 0 \\ 
0 & \cos \delta _B
\end{array}
\right) U_B^{^{\prime }}.  \label{3}
\end{equation}
The optimum is in the sense that we can choose suitable parameterizations $%
U_{A(B)}$, $U_{A(B)}^{\prime }$, $\theta _{A(B)}$, $\delta _{A(B)}$ to
realize optimal entanglement concentration. Similar as those shown for
entangled pure states, we can perform the single qubit unitary operations $%
U_{A(B)}^{\left( \prime \right) }$ and POVM by varying the SPR $A(B)i$,
therefore our linear optical protocol can implement entanglement
concentration for an arbitrary initial entangled mixed state.

It is interesting to compare our protocol with Thew and Murno's, which used
four VBS to obtain an effective transmission matrix $A\otimes B=\text{diag}
\left( \eta _{HA}\eta _{HB},\eta _{HA}\eta _{VB},\eta _{VA}\eta _{HB},\eta
_{VA}\eta _{VB}\right) $. In their protocol, Thew and Munro asked the
question why there are four individually tunable filters $\eta _{HA}$, $\eta
_{HB}$, $\eta _{VA}$, $\eta _{VB}$. As we can see from the local quantum
operator in (3), four individually tunable filters are the minimum
requirement for implementing arbitrary local quantum operations! Obviously
the effective transmission matrix in Thew and Munro's scheme can be obtained
by setting $\cos \theta _{A(B)}=\eta _{HA(B)}$ and $\cos \delta
_{A(B)}=\eta_{VA(B)}$, therefore all Thew and Munro's discussion about
entanglement concentration can be applied to ours. However, our protocol is
more feasible within current linear optical technology because it need only
Mach-Zehnder interferometer and some HWP and QWP.

In conclusion, we have proposed an experimentally feasible protocol for
implementing arbitrary local single-qubit quantum operations on individual
polarization-entangled photon pairs using linear optical devices (PBS, HWP,
QWP). Based on this technology, we have discussed concentration for
entangled pure and mixed states with a single copy. We emphasis its
simplicity and university. For example, it can also be used in multi-partite
entanglement manipulation \cite{Zhang00}. We believe the scheme should
provide a useful tool in the exploration of various quantum information
processing.

\begin{figure}[!t]
\begin{center}
\vspace*{-0.0cm}
\par
\resizebox *{8cm}{7cm}{\includegraphics*{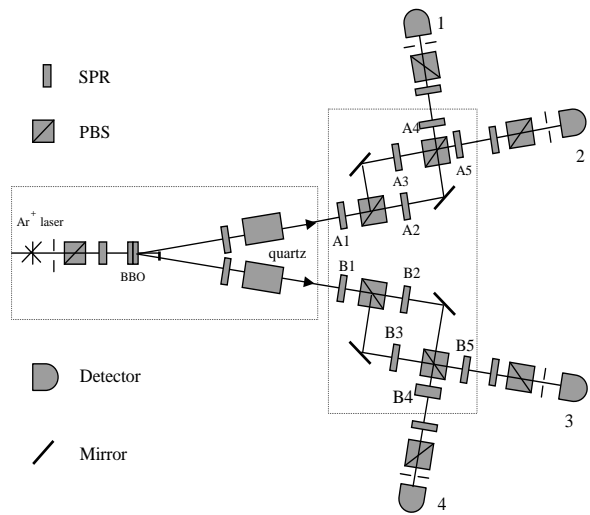}}
\end{center}
\par
\vspace*{-0.0cm}
\caption{ Experimental set-up for entanglement concentration. }
\label{fig:cir}
\end{figure}


\begin{references}

\bibitem{divin}  D.P. Divincenzo, Science {\bf 270}, 255 (1995); C.H. Bennett and D.P. Divincenzo, Nature (London) {\bf 404}, 247 (2000).
\bibitem{teleportation} C.H. Bennett, G. Brassard, C. Cr¨¦peau, R. Jozsa, A. Peres, and W.K. Wootters, Phys. Rev. Lett. {\bf 70}, 1895 (1993);
D. Bouwmeester, J.-W. Pan, K. Mattle, M. Eible, H. Weinfurther, and A. Zeilinger, Nature (London) {\bf 390}, 575 (1997);
J.-W. Pan, M. Daniell, S. Gasparoni, G. Weihs, and A. Zeilinger, Phys. Rev. Lett. {\bf 86}, 4435 (2001).
\bibitem{dense} C.H. Bennett and S.J. Wiesner, Phys. Rev. Lett. {\bf 69}, 2881 (1992);
K. Mattle, H. Weinfurter, P.G. Kwiat, and A. Zeilinger, Phys. Rev. Lett. {\bf 76}, 4656 (1996).
\bibitem{Naik00}  A.K. Ekert, Phys. Rev. Lett. {\bf 67} 661 (1991);
D.S. Naik, C.G. Peterson, A.G. White, A.J. Berglund, and P. G. Kwiat, Phys. Rev. Lett. {\bf 84}, 4733(2000);
W. Tittel, J. Brendel, H. Zbinden, and N. Gisin, Phys. Rev. Lett.  {\bf 84}, 4737 (2000).
\bibitem{Ben96} C.H. Bennett, G. Brassard, S. Popescu, B. Schumacher, J.A. Smolin, and W.K. Wootters, Phys. Rev. Lett. {\bf 76}, 722 (1996).
\bibitem{Vidal99} G. Vidal, Phys. Rev. Lett. {\bf 83}, 1046 (1999).
\bibitem{Linden98}  N. Lindel S. Massar, and S. Popescu, Phys. Rev. Lett. {\bf 81}, 3279 (1998);
A. Kent, N. Linden, and S. Massar, Phys. Rev. Lett. {\bf 83}, 2656 (1999).
\bibitem{Pan00} J.-W. Pan, C. Simon, A. Brukner, and A. Zeilinger, Nature {\bf 410} 1067 (2001); 
J.-W. Pan, S. Gasparoni, R. Ursin, G. Weihs, and A. Zeilinger, Nature {\bf 423} 417 (2003); 
T. Yamamoto, M. Koashi, A. Kayazdemir, and N. Imoto, Nature {\bf 421}, 342 (2003); 
Z. Zhao, T. Yang, Y.-A. Chen, A.-N. Zhang, J.-W. Pan, Phys. Rev. Lett. {\bf 90}, 207901 (2003).
\bibitem{Thew01}  R. T. Thew and W. J. Munro, Phys. Rev. A {\bf 63}, 030302(R) (2001); Phys. Rev. A {\bf 64} 022320 (2001).
\bibitem{Kwiat01}  P. G. Kwiat S. Barraza-Lopez, A. Stefanov, and N. Gisin, Nature {\bf 409}, 1014 (2001).
\bibitem{pop} S. Popescu, Phys. Rev. Lett. {\bf 74}, 2619 (1995).
\bibitem{sten96}  S. Stenholm, Opt. Comm. {\bf 123}, 287 (1996).
\bibitem{proof}  Any $2\times 2$ matrix $M_1$ can be diagonalized by two
unitary matrices as $M_1=U_2%
\mathop{\rm diag}%
\left( \gamma _1,\gamma _2\right) U_1$, which yields the relation $%
M_2^{^{\prime }\dagger }M_2^{\prime }=%
\mathop{\rm diag}%
\left( 1-\gamma _1^2,1-\gamma _2^2\right) $, where $M_2^{\prime
}=M_2U_1^{\dagger }$, $0\leq \gamma _i\leq 1$. We can write $M_2^{\prime
}=U_3\left( 
\begin{array}{ll}
a_1 & a_2 \\ 
0 & a_4
\end{array}
\right) $ with $U_3$ a unitary matrix. Substituting it into the relation
above, we derive $a_2=0$, $\left| a_1\right| =\sqrt{1-\gamma _1^2}$, $\left|
a_4\right| =\sqrt{1-\gamma _2^2}$, and $M_2=U_3%
\mathop{\rm diag}%
\left( \sqrt{1-\gamma _1^2},\sqrt{1-\gamma _2^2}\right) U_1$, that is, $M_i$
can be implemented by $A_1=%
\mathop{\rm diag}%
\left( \cos \theta ,\cos \vartheta \right) $, $A_2=%
\mathop{\rm diag}%
\left( \sin \theta ,\sin \vartheta \right) $ with post-selecting unitary
operations $U_2$ and $U_3$ respectively.

\bibitem{Kwiat95}  P. G. Kwiat, E. Waks, A.G. White, L. Appelbaum, and P.H. Eberhard, Phys. Rev. A {\bf 60}, 773(R) (1999).
\bibitem{White99}  A. G. White, D.F.V. James, P.H. Eberhard, and P.G. Kwiat, Phys. Rev. Lett. {\bf 83}, 3103 (1999).
\bibitem{Kwiat002}  P. G. Kwiat, A. J.Berglund, J. B. Altepeter, and A. G. White, Science {\bf 290}, 498 (2000);
Y.-S. Zhang, Y.-F. Huang, C.-F. Li, and G.-C. Guo, Phys. Rev. A ${\bf 66}$, 062315 (2002); 
C. Zhang, Phys. Rev. A {\bf 69}, 014304 (2004).
\bibitem{Schl95}  J. Schlienz and G. Mahler, Phys. Rev. A {\bf 52}, 4396 (1995).
\bibitem{Vers00}  F. Verstraete J. Dehaene, and B.D. Moor, Phys. Rev. A {\bf 64}, 010101(R) (2001).
\bibitem{Horo98}  P. Horodecki, M. Horodecki, and R. Horodecki, Phys. Rev. A {\bf 60} 1888 (1999).
\bibitem{Zhang00}  C.-W. Zhang, C.-F. Li, Z.-Y. Wang, and G.-C. Guo, Phys. Rev. A {\bf 62}, 042302 (2000).
\end{references}
\end{document}